\documentstyle[aps,twocolumn]{revtex}
\begin{document}
\title{Fragmented and Single Condensate Ground
States of Spin-1 Bose Gas}
\author{Tin-Lun Ho$^{+}$ and Sung Kit Yip$^{\ast}$}
\address{$^{+}$Department of Physics,  The Ohio State
University, Columbus, Ohio 43210\\
$^{\ast}$Physics Division, National Center for Theoretical
Sciences, P. O. Box 2-131, Hsinchu, Taiwan 300}

\maketitle

\begin{abstract}
We show that the ground state of a spin-1 Bose gas with an antiferromagnetic 
interaction is a fragmented condensate in uniform magnetic fields. 
The number fluctuations in each spin component change rapidly from being
enormous (order $N$) to exceedingly small (order 1) 
as the magnetization of 
the system increases. A fragmented condensate can be turned into 
single condensate state by magnetic field gradients. The conditions for
existence and method of detecting fragmented states are presented. 
\end{abstract}

\vspace{0.2in}

Bose-Einstein condensation (BEC) was first introduced as a phenomenon 
in non-interacting Bose systems. The concept of BEC was generalized to 
interacting Bose systems in 1956 by Penrose and Onsager\cite{Penrose}. 
A system of $N$ Boson is considered BE condensed if its single particle 
density matrix has one and only
one macroscopic eigenvalue (i.e. of order $N$). The corresponding
eigenfunction is identified as the quantum state macroscopically occupied. 
This characterization is in good agreement with the recent BEC 
experiments on magnetically trapped alkali atoms, 
which are effectively scalar Bosons since their spins are frozen. 
Recently, optical trapping of Bose condensate has become 
possible\cite{optical}. An optical trap confines all spin states, and 
the nature of the condensate depends on the magnetic
interaction\cite{Ho}\cite{Jap}.  In the case of $^{23}$Na where the 
interaction is antiferromagnetic\cite{Ho}, the single 
condensate interpretation appears to agree with 
experiments\cite{Stenger}. Since $^{23}$Na is a spin-1 Boson, the 
macroscopically occupied state is a three component spinor. 

However, in a recent paper, Law, Pu, and Bigelow\cite{Law} have 
pointed out that the Hamiltonian in ref.\cite{Ho}\cite{Jap} in zero magnetic 
field has a singlet ground state, 
with properties drastically different from those of 
spinor condensates. They, however, did not discuss how their results will
reconcile with the MIT experiment\cite{Stenger}.  The spin singlet turns out 
to be a ``fragmented" condensate, meaning that its density matrix has more
than one macroscopic eigenvalue.  The possibility of fragmented condensate 
was first discussed by Nozieres and Saint James (NS) \cite{NS}, who 
concluded that it cannot occur in homogenous scalar Bose gas with repulsive
interactions. 

As we shall see, the singlet state is unstable. It quickly gives 
way to a much more
generic fragmented state (called coherent-fragmented state) as 
magnetization increases. The coherent-fragmented state has identical occupation
number as the single condensate state in each spin component, except that 
there is no phase coherence between different spin components. 
As a result, these two states 
cannot be distinguished by density measurements such as the MIT 
experiment\cite{Stenger}. 
To tell them apart, it is necessary to measure the phase coherence between
different spin components, which can be done by performing 
Stern-Gerlach experiments along different axes. 
The origin of the fragmented state turns out to spin conservation.  
As a result, magnetic field gradients which destroy spin conservation 
can deform fragmented states toward single condensate states. The degree 
of deformation depends on particle number and the strength of 
the field gradient. 
In the following, we first
consider homogenous Bose gas, where ``fragmentation" can be 
discussed most efficiently. 
Discussions on trapped gases will follow. 

{\bf I. Homogenous Spin-1 Bose Gas :}
Consider a spin-1 Bose gas with Hamiltonian\cite{Ho}\cite{Jap}
$\hat{H} = \hat{h} + \hat{V}$, $\hat{h}=\int \psi^{\dagger}_{\mu}
h({\bf x})_{\mu\nu}  \psi^{}_{\nu}$, $h({\bf x})_{\mu\nu}\equiv
[-\frac{\hbar^2}{2M}{\bf \nabla}^2 + U({\bf x})]\delta_{\mu\nu}
- \gamma {\bf B}\cdot {\bf S}_{\mu\nu}$,  $\hat{V} = \frac{1}{2}\int
\psi^{\dagger}_{\alpha}\psi^{\dagger}_{\mu}
\psi^{}_{\nu}\psi^{}_{\beta}\left[ c_{0}\delta_{\alpha\beta}
+ c_{2} {\bf S}_{\alpha\beta}\cdot{\bf S}_{\mu\nu} \right]$, where 
$c_{0}, c_{2}>0$,  
$M$ and $\gamma$ are the mass and gyromagnetic ratio of the Boson 
respectively, and ${\bf B}$ is a uniform magnetic field. The field 
operator $\psi_{\mu}({\bf x})$ ($\mu = \pm1, 0$) can be expanded as 
$\phi_{\mu}({\bf x})$$=\Omega^{-1/2}$
$\sum^{}_{{\bf k}\neq 0}$$e^{i{\bf k}\cdot {\bf x}} a^{}_{\mu}({\bf k})$
where $\Omega$ is the volume of the system.  For simplicity, we shall 
denote $a^{}_{\mu} ({\bf k}=0)$ simply as $a^{}_{\mu}$. 
Denoting the part of 
$\hat{H}$ containing $a_{\mu}$ alone as $\hat{H}_{o}$ and the rest as 
$\hat{H}_{ex}$, ($\hat{H} = \hat{H}_{o} + \hat{H}_{ex}$), we have 
\begin{equation}
\hat{H}_{o} = \frac{c_{2}}{2\Omega}\left( {\bf S}^{2} - 2N \right) 
- \gamma {\bf B}\cdot {\bf S} + C
\label{Ho} \end{equation} 
where ${\bf S}$$=a^{\dagger}_{\mu}$${\bf S}^{}_{\mu\nu}$
$a^{}_{\nu}$, $N$$= a^{\dagger}_{\mu}a^{}_{\mu}$, 
and $C$$= \frac{c_{o}}{2\Omega}$$[ N^{2} - N]$. 

To find the ground state $|G\rangle$, we first find the ground state of 
$\hat{H}_{0}$ (denoted as $|F\rangle$) and then study 
condensate depletion effects 
due to $\hat{H}_{ex}$.  From eq.(\ref{Ho}), we see that 
$|F\rangle= |S^{total}=S; S^{total}_{z}=S\rangle$, where 
$S$ is the integer closest to $(\gamma B \Omega /c_{2})-\frac{1}{2}$, which is 
the minimum of $\langle H_{o}\rangle_{F} =
(c_{2}/2\Omega)S(S+1) - \gamma B S - c_{2} \frac{N}{\Omega} + C $. 
The equilibrium magnetization is 
$s_{o} = S/N = \gamma B/c_{2}n + 
0(N^{-1})$ and $n = N/\Omega$.  In the following, we shall only consider 
the case $\gamma B< c_{2}n $, where $S/N$ ranges from 0 to 1. 

In contrast, the optimum single condensate state in a magnetic field is 
\begin{equation}
|SC\rangle = \frac{1}{\sqrt{N!}} \left( 
\sqrt{\frac{N_{1}}{N}} a^{\dagger}_{1} + e^{i\chi}
\sqrt{ \frac{N_{-1}}{N}} a^{\dagger}_{-1} 
\right)^{N}|{\rm vac}\rangle, 
\label{SC} \end{equation}
where $\chi$ is a relative phase, $N_{\pm 1} = N(1\pm y)/2$, and 
$y$ is the magnetization. 
(We shall from now on absorb $\chi$ into the operators for simplicity. 
The importance of $\chi$ will be discussed at the end.)
The equilibrium magnetization $y_{o}$ is obtained by 
minimizing the energy 
$\langle \hat{H}_{o}
\rangle_{SC}= \frac{c_{2}}{2\Omega}N(N-1)y^2 - \gamma B N y + C 
\equiv G(y)$, and is 
$y_{o} =\gamma B\Omega/[c_{2}(N-1)]= \gamma B/gn + 0(N^{-1})$. 
The energy difference between $|F\rangle$ and $|SC\rangle$ is
$\Delta E = \langle {\cal H}_{o} \rangle_{_{SC}}$$-
\langle {\cal H}_{o} \rangle_{_{F}}$
$= G(y_{o}) - G(s_{o}) + (c_{2}N/2\Omega)(2-s_{o} - s^2_{o})$. 
Since $s_{o}<1$, the last term in $\Delta E$ is positive. Moreover, 
$y_{o}-s_{o}\sim 0(N^{-1})$, we have $G(y_{o}) - G(s_{o})
= \frac{c_{2}}{2}\frac{N(N-1)}{\Omega}(y_{o} - s_{o})^2 \sim 
\frac{c_{2}}{2\Omega}$, which is smaller than the last term in $\Delta E$
by a factor of $N$ and can therefore be ignored. This shows that 
{\em for a homogenous Bose gas}, the states $|F\rangle$ and $|SC\rangle$
are degenerate in the thermodynamic limit
($N\rightarrow \infty, \Omega \rightarrow \infty, N/\Omega \rightarrow$
finite), since their energies are of order $N$ but their difference 
($c_{2}n$) is of order 1.  The relative stability between 
$|F\rangle$ over $|SC\rangle$ is therefore very delicate. To discuss
this stability, it is necessary to understand the structure of 
$|F\rangle$, which turns out to be very remarkable. 

{\bf II. Super- and Coherent- Fragmentation :} A simple exercise shows that 
$\Theta^{\dagger} \equiv -2a_{1}^{\dagger}a^{\dagger}_{-1}+a^{\dagger 2}_{0}$ 
creates a singlet pair of spin-1 Bosons.  The ground state $|F\rangle = |S;
S\rangle$ is therefore given by 
\begin{equation}|S;S\rangle = \frac{1}{\sqrt{f(Q;S)}} a^{\dagger S}_{1} 
\Theta_{}^{\dagger Q}|{\rm vac}\rangle , \,\,\,\,\,\,\,\, Q = (N-S)/2, 
\label{SS}
\end{equation}
where $f(Q;S)$ is the normalization constant\cite{derivation}
\begin{equation}
f(Q; S) = S! Q! 2^{Q} \frac{(2Q + 2S + 1)!!}{ (2S +1)!!}. 
\label{norm} \end{equation}
Using eq.(\ref{norm}),
it is easy to show that the single particle density matrix of $|F\rangle$ 
is diagonal, $(\hat{\rho}^{F})_{\alpha\beta} = 
\langle a_{\beta}^{\dagger}a^{}_{\alpha} \rangle_{F}$
$= N_{\alpha} \delta_{\alpha\beta}$, with 
\begin{eqnarray}
N_{1} = \frac{ N(S+1) + S(S+2) }{2S + 3}, \,\,\,\,\,
N_{-1} = \frac{ (N-S)(S+1) }{2S + 3},  \nonumber \\
N_{0}  = \frac{N-S}{2S+3}.  \label{number}
\end{eqnarray}
Since $\hat{\rho}^{F}$ has more than 
one macroscopic eigenvalue, $|F\rangle$ is a 
{\em fragmented} condensate for all $S<N$\cite{NS}.  Again using
eq.(\ref{norm}), the squared number fluctuation 
$(\Delta \hat{N}_{1})^2$$\equiv$
$\langle (a^{\dagger}_{1} a^{}_{1} - \langle a^{\dagger}_{1} a^{}_{1}
\rangle)^2 \rangle$ for spin $\mu = 1$  can be shown to be 
\begin{eqnarray}
(\Delta \hat{N}_{1})^{2}
= & \left(\frac{N}{2S+3}\right)^2\left( \frac{S+1}{2S + 5} \right)
+ \left(\frac{3N}{(2S+3)^2}\right) \left( \frac{S+1}{2S+5} \right)
\nonumber \\
 & + \left(\frac{S+1}{2S+5}\right) \left( \frac{S^2 -3S}{(2S + 3)^2}\right).
\label{numberflu} \end{eqnarray}
Moreover, we have $(\Delta \hat{N}_{-1})^2$
$= (\Delta \hat{N}_{1})^2$$=(\Delta \hat{N}_{0})^2/4$ from the 
relations $\hat{N}_{-1}$$= \hat{N}_{1}-S$, and $\hat{N}_{0}$
$= N + S  - 2\hat{N}_{1}$. 

Eqs.(\ref{number}) and (\ref{numberflu}) show that when $S=0$, the system
has $N_{1}=N_{0}=N_{-1}= N/3$\cite{Law} with enormous 
fluctuations $\Delta N_{\alpha}\sim N$.
On the other hand, both $N_{0}$ and $\Delta N_{\alpha}$ shrink rapidly as 
$S$ increases. When $S$ becomes macroscopic, 
$N_{0}$ and $\Delta N_{\alpha}$ become order 1, whereas 
$N_{\pm 1}$ remains macroscopic, $N_{\pm 1} \rightarrow (N\pm S)/2$. 
The exceedingly small fluctuations $\Delta N_{\alpha}$ means that state
$|S;S\rangle$ can be well approximated by 
\begin{equation}
|S; S\rangle \sim |N_{1}, N_{-1}\rangle \equiv \frac{a^{\dagger N_{1}}_{1}
a^{\dagger N_{-1}}_{-1}}{\sqrt{N_{1}! N_{-1}!}}|{\rm vac}\rangle . 
\label{excellent} \end{equation}  
To distinguish different fragmented states, we call those with 
$(\Delta \hat{N}_{\alpha})\sim N$ ``super"-fragmented states, 
and those with $(\Delta \hat{N}_{\alpha})\sim 1$ ``coherent"-fragmented state. 
In the thermodynamic limit, super-fragmented is a singularity which occurs only 
at $S/N=0$. All other states with $S/N\neq 0$ are coherent-fragmented. 

The origin of the small fluctuations in the coherent-fragmented state 
can be understood as follows.  Noting that 
$S^{}_{+}$$=\sqrt{2}(a^{\dagger}_{1}a^{}_{0}$$+a^{\dagger}_{0}a^{}_{-1})$, 
eq.(\ref{Ho}) can be written as $\hat{H}_{o}$$=\hat{H}_{A}$$+\hat{H}_{B}$, 
\begin{equation}
\hat{H}_{A} = \frac{c_{2}}{2\Omega}\left[ S^{2}_{z} + 
(2N^{}_{0}-1)(N^{}_{1}+N^{}_{-1}) \right] - \gamma B S_{z}, \label{HA}
\end{equation}
and
$\hat{H}_{B}$$=\frac{c_{2}}{\Omega}$$(a^{\dagger 2}_{0}a^{}_{1}a^{}_{-1}$
$+ h.c.)$. 
 $\hat{H}_{B}$ contains all terms which are responsible for
 transformation among spin species.
In $\hat H_A$ we have dropped terms depending only on $N$.
Clearly, $\hat{H}_{A}$ is minimized when $N_{0}=0$. If $|N_{1}, N_{-1},
N_{0}\rangle$ denotes the state with $N_{\mu}$ Bosons with spin $\mu$, 
the ground state of $\hat{H}_{A}$ is 
$|N^{}_{1}, N^{}_{-1}, 0\rangle$$=|N^{}_{1}, N^{}_{-1}\rangle$, where 
$N_{\pm 1}= (N\pm S)/2$, and $S = \gamma B\Omega /c_{2}$. The ground state
energy is $E_{o} = - c^{2}S^2/(2\Omega)$.
The effect of $\hat{H}_{B}$ is to mix in $N_{o}\neq 0$ states 
$|q \rangle = |N^{}_{1} -q, N^{}_{-1} - q, 2q\rangle$. To 
leading order in $N$, we have  $\hat{H}_{o} - E_{o} 1 = \tilde{H}$, 
\begin{equation}
\tilde{H} = \frac{c_{2}N}{\Omega} 
\sum_{q=0, 1, 2, ..} \left[ 2q |q\rangle \langle q|
+ {\lambda_{q}}(|q+1\rangle \langle q| + h.c. ) \right]
\label{mixing} \end{equation}
where $\lambda_{q}=$$\sqrt{(N_{1}-q)(N_{-1}-q)(2q + 2)(2q+1)}/N$
For $1<<q<<N_{1}, N_{-1}$, $\lambda_{q}$$=x(q+$$\frac{3}{4})$, 
$x=$$\sqrt{1-(S/N)^2}$. To estimate the energy  of eq.(\ref{mixing}), 
we consider the following variational state 
$|\Psi\rangle = \sqrt{\alpha} \sum_{q}(-1)^{q}e^{-\alpha q/2}$ with 
energy $E = $$\frac{c_{2}N}{\Omega}$$(\frac{1}{\alpha} $$- x e^{-\alpha/2}$
$[\frac{1}{\alpha}$$+ \frac{3}{4}])$. The minimum condition
is $e^{\alpha/2}$$= x(1+\frac{1}{2}\alpha$$ + \frac{3}{2}\alpha^2)$. To leading
oder in $S/N$, we have $\alpha = \sqrt{2}(S/N)$. The energy correction 
$\langle \tilde{H} \rangle_{\Psi}$ is
therefore of order $\frac{c_{2}N}{\Omega}$, which is lower than $E_{o}$ by a 
factor of $N$. Thus, to the leading order in $N$,
$\hat{H}_{o}$ can be replaced by $\hat{H}_{A}$ [eq.(\ref{HA})] with $N_{0}=0$, 
with ground state eq.(\ref{excellent}). 

For later discussions, it is useful to compare the coherent-fragmented state 
eq.(\ref{excellent}) with the single condensate state $|SC\rangle$
eq.(\ref{SC}). Writing 
$|SC\rangle$$=\sum_{\ell = -N_1}^{N_{-1}}$$\left( 
\frac{ N_{1}^{N_{1}+\ell}N_{-1}^{N_{-1}-\ell} }
{(N_{1}+\ell)! (N_{-1} - \ell)!} \frac{N!}{N^{N}} \right)^{1/2}
|\ell \rangle$, where $|\ell\rangle \equiv |N_{1}+\ell, N_{-1}-\ell\rangle$, 
and using the Stirling formula, it is straightforward to show
that the single condensate is a Gaussian sum of coherent fragmented states
\begin{equation}
|SC\rangle \stackrel{\sim}{=}
 (\pi \sigma^{2})^{-1/4}\sum_{\ell = -N_{1}}^{N_{-1}}
e^{-\ell^2/2\sigma^2}e^{\xi \ell}|\ell\rangle
\label{sum} \end{equation}
where $\sigma^2 = 2N_{1}N_{-1}/N$, $\xi = (N_{-1}^{-1} -
N_{1}^{-1})/4$\cite{CommentA}. 
Within the space of $\mu = \pm 1$, the density matrix of 
$|SC\rangle$ is 
\begin{equation}
\hat{\rho}^{SC}_{\mu\nu} \equiv \langle a^{\dagger}_{\nu}a^{}_{\mu}\rangle_{SC}
= \left( \begin{array}{cc} N_{1} & \sqrt{N_{1}N_{-1}} \\
\sqrt{N_{1}N_{-1}} & N_{-1} \end{array}   \right) , 
\label{dmsc} \end{equation}
The off-diagonal elements $\sqrt{N_{1}N_{-1}}$ are absent in
$\hat{\rho}^{F}$.

{\bf III.  Persistence of Fragmentation :} So far, we have ignored 
condensate depletion (i.e. due to $\hat{H}_{ex}$), and various realistic 
atomic physics effects, (see below). In the following, we shall focus on 
coherent-fragmented states as they are most common.
Following the method of Huang and Yang\cite{LeeYang} to extract the dominate
condensate depletion effect from $\hat{H}_{ex}$, we find that the 
effective Hamiltonian for condensate depletion for both 
single condensate and fragmented state have identical Bogoliubov form, 
which is obtained by replacing the operators 
$a^{\dagger}_{\pm 1}({\bf k}= {\bf 0})$ by the c-number $\sqrt{N_{\pm 1}}$. 
This means that the thermodynamic degeneracy of 
$|F\rangle$ and $|SC\rangle$ cannot be lifted by condensate depletion, 
as it gives rise to same energy change in both cases. 
(This result also applies to the trapped Bosons, in which case
the states (${\bf k}$, $-{\bf k}$) are replaced by opposite angular momentum
states.)
As for atomic physics effects such as mixing of different hyperfine states 
and quadratic Zeeman effects, they all respect spin rotational symmetry along
$z$\cite{quadratic}. 
The new ground state $|G\rangle$ with all these effects included 
therefore remains an eigenstate of $S_{z}$, which implies 
$\langle  G|a^{\dagger}_{1} a^{}_{-1}|G \rangle =0$
as $a^{\dagger}_{1}a^{}_{-1}$ changes $S^{(o)}_{z}$.
Thus, these effects cannot reassemble a fragmented
state into a single condensate state, for 
$\langle a^{\dagger}_{1} a^{}_{-1} \rangle_{SC}= \sqrt{N_{1}N_{-1}}$ 
is of order $\sim N$ instead of 0. 

{\bf IV. Reassembling Fragmented States into Single Condensates }: 
Since spin conservation protects fragmentation, 
we consider perturbations that break spin symmetry. 
The natural candidate is a magnetic field gradient $G'$.
To be concrete, we take ${\bf B}({\bf x}) = B_{o}(\hat{\bf z} +
G'[x\hat{\bf x} - z \hat{\bf z}])$.  The condition of small field gradient is
that $G'\Omega^{-1/3}<<1$. It is useful to choose local field direction
$\hat{\bf B}({\bf r})$ as the spin quantization axis. This is done by
performing a unitary transformation
$\hat{U}= \prod_{i=1}^{N} e^{-i{\mbox{\boldmath $\theta$}}({\bf x}_{i})\cdot
{\bf S}_{i}}$, where 
${\mbox{\boldmath $\theta$}} = \hat{\bf z}\times \hat{\bf B} = G'x\hat{\bf y}
+ 0(G'^2)$. The interaction $\hat{V}$ is 
invariant under $\hat{U}$ because it is a spin conserving contact interaction. 
However, $\hat{U}^{\dagger} {\bf p} \hat{U}
= {\bf p} + \hbar G' S_{y} \hat{\bf x} + 0(G'^3)$. This causes $\hat{H}_{o}$ 
$\rightarrow (\hat{U}^{\dagger}\hat{H}_{o}\hat{U})_{o} = 
\hat{H}_{o} + \hat{H}_{1} $, where
$\hat{H}_{1}$$= \epsilon$$\sum_{i=1}^{N} (S^{y}_{i})^2$ and 
$\epsilon \equiv \frac{\hbar^2 G'^{2}}{2M}$. When operated on the coherent 
fragmented states eq.(\ref{excellent}), 
$\hat{H}_{1}$ is reduced to 
$\hat{H}_{1} = -(\epsilon/2) ( a^{\dagger}_{1}a^{}_{-1} +
a^{\dagger}_{-1}a^{}_{1}) + \epsilon N/2$. 
The effect of $\hat{H}_{1}$ on the state $|N_{1},
N_{-1}\rangle$ is to generate the set $\{ |\ell\rangle \equiv |N^{}_{1}+\ell, 
N^{}_{-1} - \ell \rangle \}$. 
Within this set, $\hat{H}'_{o}$ has the tight-binding form
\begin{equation}
\hat{H'}_{o} = \sum_{\ell = 0, \pm 1, ...}  \left[ \frac{2c_{2}\ell^2}{\Omega}
|\ell\rangle \langle \ell |  - 
\frac{t_{\ell}}{2} 
\left(  |\ell +1 \rangle \langle \ell | + h.c.   \right) \right]
\label{tight} \end{equation}
where $t_{\ell} \equiv \epsilon 
\sqrt{(N_{1} + \ell + 1)(N_{-1}-\ell)}$.  The eigenstates of 
eq.(\ref{tight}), $|\Psi\rangle = \sum_{\ell}\Psi_{\ell}|\ell\rangle$,
satisfies 
the Schr\"odinger equation $E\Psi_{\ell} = (2c_{2}/\Omega)\ell^2 \Psi_{\ell}
- (t_{\ell}\Psi_{\ell +1} + t_{\ell -1}\Psi_{\ell -1})/2$. 
Although this equation can be solved numerically, it is more illuminating to 
consider the following analytic approximation, which turns out to be very
accurate. As we shall verify later, the number of $\ell$ terms in the ground
state $|\Psi\rangle$ is much less than the typical value of $N_{1}$ and
$N_{-1}$, so that we can replace $(t_{\ell}+ t_{\ell -1})/2$
$\sim$$\epsilon \sqrt{N_{1}N_{-1}}$, and 
$(t_{\ell}-t_{\ell -1})/2$$\sim$$-\epsilon \sqrt{N_{1}N_{-1}}
\xi$, where $\xi = (N_{-1}^{-1}-N_{1}^{-1})/4$ as defined before. 
The Schr\"odinger equation in the continuum limit is then 
$( \frac{\Omega E}{4c_{2}}$$+ \eta^4)\Psi_{\ell}$$= - \frac{1}{2} \eta^4$
$\frac{ {\rm d}^2\Psi_{\ell} }{{\rm d} \ell^2}$
$ + \eta^4 \xi \frac{ {\rm d}\Psi_{\ell} }{{\rm d}\ell}$
$+\frac{1}{2} \ell^2$$\Psi_{\ell}$, where 
\begin{equation}
\eta^4 \equiv \epsilon \sqrt{N_{1}N_{-1}} \Omega/(4c_{2}).
\label{eta}  \end{equation}
The normalized ground state is 
\begin{equation}
|\Psi\rangle = (\pi \eta^2)^{-1/4}\sum_{\ell}
e^{-\ell^2/2\eta^2}e^{\xi \ell}|\ell\rangle, 
\label{newground} \end{equation}
with a density matrix 
$(\hat{\rho}_{\Psi})_{\alpha\beta} = \langle \Psi|
a^{\dagger}_{\alpha}a^{}_{\beta}|\Psi\rangle$ ($\alpha, \beta = \pm 1)$, 
\begin{equation}
(\hat{\rho}_{\Psi})_{\alpha\beta} = 
\left( \begin{array}{cc} 
N_{1}  & \sqrt{N_{1}N_{-1}} e^{-1/4\eta^2} \\
\sqrt{N_{1}N_{-1}} e^{-1/4\eta^2} & N_{-1} \end{array} \right). 
\end{equation}
The eigenvalues of $\hat{\rho}_{\Psi}$ are 
\begin{equation}
\lambda_{\pm}=\frac{1}{2}\left[ N \pm \sqrt{ 
N^2 e^{-1/2\eta^2} + S^2 ( 1- e^{-1/2\eta^2}) } \right].
\label{eigen} \end{equation}
For zero field gradient, $\eta \rightarrow 0$, $\lambda_{\pm} \rightarrow 
\frac{1}{2}(N\pm S)$. For large field gradients, 
$\eta >>1$, eq.(\ref{newground}) reduces to eq.(\ref{SC}), and 
$\lambda_{+} \rightarrow N$, and $\lambda_{-} \rightarrow 0$.
The systems turns into a single condensate. Note that even for $\eta
\sim 5$, the system is essentially a single condensate state. 
Using the expression of $c_{2}$ in ref.\cite{Ho}, $c_{2} 
= 4\pi \hbar^2 \Delta a_{sc}/M$, 
$\Delta a_{sc} = (a_{2} - a_{0})/3$, we have 
$\eta\equiv$$\eta_{o}(1-(\frac{S}{N}))^{1/8}$, and 
$\eta_{o}$$= \left[\epsilon \Omega N/(4c_{2}) \right]^{1/4}$
$ = \left[ \frac{ (G'\Omega^{1/3})^2 N^{4/3}}
{ 32\pi (\Delta a_{sc} n^{1/3}) }\right]^{1/4}$. 
Since $\eta \sim N^{1/3}$, arbitrarily small field gradients will change a
fragmented state into a single condensate for homogenous Bose gas in the
thermodynamic limit. 

However, trapped Bose gases with $N\stackrel{<}{\sim} 10^6$ are in mesoscopic
rather than thermodynamic limit. In this limit, 
super- and coherent-fragmented states are no longer singularities 
in the phase space $(\frac{S}{N}, \epsilon)$. 
Instead, they occupy a finite region in the phase space which crosses over to
single condensate states in a continuous manner. This implies the possibility
of observing fragmented condensates in trapped gases.

{\bf V. Trapped Spin-1 Bose Gas :} When ${\bf B}\neq 0$, different spin 
components have different spatial extents. For cylindrical traps, we write  
$\psi_{\mu}({\bf x}) = f^{}_{\mu}({\bf x}) 
a^{}_{\mu} + \phi_{\mu}({\bf x})$, where $f^{}_{\mu}$'s are normalized 
and cylindrically symmetric wavefunction of the condensate with spin $\mu$, 
and $\phi^{}_{\mu}$ is the non-condensate part of $\psi_{\mu}$
satisfing $\int f_{\mu}\phi =0$. 
The exact forms of $f_{\mu}$ are determined by energy
minimization.  As in the homogenous case, we write 
$\hat{H} = \hat{H}_{o}+ \hat{H}_{ex}$ where $\hat{H}_{o}$ contains only
$a^{}_{\mu}$.  $\hat{H}_{o}$ can again be written as 
$\hat{H}_{A}$$+\hat{H}_{B}$, where $\hat{H}_{A}$ contains only $S_{z}$, 
$N^{}_{1}+N^{}_{-1}$, $N^{}_{0}$; 
and $\hat{H}_{B}$ has the same form as before 
 but with a different coefficient.
$\hat H_A$ is again minimized at $N_{0}=0$.
The effect of $\hat H_B$ is again to give rise to 
eq.(\ref{mixing}) with a weaker hopping term. \cite{LDA}
Our previous results therefore apply, i.e. 
to leading order in $N$, $\hat{H}_{B}$ can be ignored and the ground state 
with macroscopic magnetization is accurately given by 
$|F\rangle$$=|N^{}_{1}, N^{}_{-1}\rangle$, $N_{\pm 1}= (N\pm S)/2$. 
The functions $f_{1}$ and $f_{-1}$ are determined
by minimizing $E= \langle \hat{H}_{o}\rangle_{F}$.
In the presence of a field
gradient, a repeat of our previous calculation shows that the Hamiltonian
$\hat{H}_{o}'$ is again given by eq.(\ref{tight}), with 
$\Omega^{-1} \rightarrow \frac{1}{4}\int \left[ (f_{1}^2 + f_{-1}^2)^2 
+ \frac{c_{o}}{c_{2}}(f_{1}^2 - f_{-1}^2)^2 \right]$, and $\epsilon \rightarrow 
\epsilon \int f_{1}f_{-1}$.  

To estimate the field gradients $G'$ and 
magnetization $S/N$ needed for observing fragmented states, 
we recall from eq.(\ref{eta}) that 
$\eta^4 = \frac{N\Omega G'^{2}}{32\pi \Delta a_{sc}} 
(1-(\frac{S}{N})^2)^{1/2}$, where we have used 
$c_{2} = 4\pi \hbar^2 \Delta a_{sc}/M$.  Next, we note that 
the maximum and minimum value of $\lambda_{+}$
in eq.(\ref{eigen}) is $N$ and $(N+S)/2$. Taking the mean of these values
$\lambda^{\ast} = (N + (N+S)/2)/2$ as a specification of the crossover
region between single condensate and fragmented state, 
we find that fragmented states with 
$\lambda_{+}<\lambda^{\ast}$ emerges when 
$G'< \left(\frac{8\pi \Delta a_{sc}}{\sqrt{N_{1}N_{-1}} \Omega 
[\int f_{1}f_{-1} ]}\right)^{1/2}/$
$\left({\rm ln} \frac{4(N+S)}{N+3S}\right)$.  For a $^{23}$Na gas ($\Delta
a_{sc} \sim 10^{-8}{\rm cm}$) with $N=10^6$, $S/N=0.2$, 
we have $\lambda^{\ast}/N=0.8$. Since $\Omega$ is
roughly the volume of the system, if we take for $\Omega \int f_{1}f_{-1}
\sim 10^{-12}$cm$^3$, we have $\lambda <\lambda^{\ast}$ when
$G'<0.46$cm$^{-1}$)\cite{crossover}.

Finally, we note that coherent-fragmented states cross over to 
super-fragmented states as $S$ decreases. 
Setting $\Delta
N_{\alpha}\sim \sqrt{N}$ as the crossover estimate from super- to
coherent- fragmentation. Eq.(\ref{numberflu})
implies that (for $N, S >>1$) 
super-fragmentation 
emerges when $S/N < 1/\sqrt{8N}$. For $N=1800$, it means $S/N < 0.016$.
Thus, super-fragmentation can only be achieved when $S/N$
is very close to zero. 
On the other hand, coherent 
fragmented states are more generic and are easier to realize. 

{\bf VI.  Distinguishing fragmented states from single condensate states: } 
Because of the difference in density matrices, the states $|SC\rangle$ and
$|F\rangle$$=|S;S\rangle$ have different rotational properties. 
Under a spin rotation 
(with Euler angles $(\alpha, \beta, \gamma)$), the density
matrix $\hat{\rho}_{\mu\nu} = \langle a^{\dagger}_{\nu}a^{}_{\mu} \rangle$
becomes $D\hat{\rho} D^{\dagger}$, where $D(\alpha, \beta, \gamma)$
is the rotational matrix. The number of spin-$\mu$ particles changes from 
$N_{\mu}= \hat{\rho}_{\mu\mu}$ to 
$N_{\mu}' = (D\hat{\rho}D^{\dagger})_{\mu\mu}$. 
For single condensate state of the form eq.(\ref{SC}) before rotation, we have 
\begin{eqnarray}
N^{'}_{\pm 1} = c^4 N^{}_{\pm 1} + s^4 N_{\mp 1} + 2c^2 s^2 \sqrt{N_{1}N_{-1}}
{\rm cos}2\tilde{\chi}   \label{scpm1} \\
N^{'}_{0} = 2s^2c^2( N^{}_{1} + N^{}_{-1} ) 
- 4c^2 s^2 \sqrt{N_{1}N_{-1}} {\rm cos}2\tilde{\chi}  \label{sc0}
\end{eqnarray}
where $\tilde{\chi}$$=\gamma - \chi/2$, 
$s={\rm sin}\beta$, $c={\rm cos}\beta$, and $\beta$ is the angle 
between the original and the new quantization axis. 

For fragmented states with occupations numbers $N_{1}, N_{0}, N_{-1}$ before
rotation, we have
\begin{eqnarray}
N^{'}_{\pm 1} = c^4 N_{\pm 1} + s^4 N_{\mp 1} + 2s^2c^2 N_{0} \label{fracpm1}\\
N^{'}_{0} = 2s^2c^2 ( N_{1} + N_{-1}) + N_{0}(c^2 - s^2)^2  \label{frag0}
\end{eqnarray}
In the coherent-fragmented regime where $N_{\pm 1}>>N_{0}$, eqs.(\ref{scpm1}) to
(\ref{frag0}) show that the difference in $N_{\mu}'$ between single condensate
state and fragmented state is the phase coherence term  $\propto 
{\rm sin}^{2}\beta \sqrt{N_{1}N_{-1}}{\rm cos} 2\tilde{\chi}$.
This difference can be detected by first preparing a sample with finite
magnetization along a specified axis and then performing a sequence of 
Stern-Gerlach experiments along a different axis, as well as different axes.  

This work is supported by a Grant
from NASA NAG8-1441, and find the NSF Grants DMR-9705295 and  DMR-9807284.


\begin{references}
\bibitem{Penrose} O. Penrose and L. Onsager, Phys. Rev. {\bf 104}, 576 (1956). 

\bibitem{optical} D. Kurn, et.al Phys. Rev. Lett. {\bf 80}, 2027 (1998).

\bibitem{Ho} T.L. Ho, Phys. Rev. Lett. {\bf 81}, 742, (1998).

\bibitem{Jap} T. Ohmi and K. Machida, J. Phys. Soc. Jpn, 
{\bf 67}, 1822 (1998).

\bibitem{Stenger} J. Stenger, et.al., {\em Nature} {\bf 396}, 345 (1998).

\bibitem{Law} C.K. Law, H. Pu, and N.P. Bigelow, Phys. Rev.  Lett. 
{\bf 81}, 5257, 1998.

\bibitem{derivation} To derive eq.(\ref{norm}), 
We define $A_{x}=(-a_{1}+a_{-1})/\sqrt{2}$, $A_{y}=-i(a_{1}+a_{-1})/\sqrt{2}$, 
$A_{z} = a_{0}$. We then have ${\bf A}^2 = \Theta$. 
Writing $|Q, S\}$$=a^{\dagger S}_{1}\Theta^{\dagger Q}|{\rm vac}\rangle$ as
$|Q, S\}$$= [ T^{Q}_{S}({\bf x})$ $e^{ {\bf A}^{\dagger}\cdot {\bf x} }$
$|{\rm vac}\rangle ]_{o}$, where 
$T^{Q}_{S}({\bf x})$$\equiv$$ (-2^{1/2})^{S}$
$(\partial_{x} + i \partial_{y})^{S}$$({\bf \nabla}^{2})^{Q}$, and $(...)_{o}$
means ${\bf x}=0$, and noting that 
$\langle {\rm vac}|$$e^{{\bf A}\cdot {\bf x}}$
$e^{{\bf A}^{\dagger}\cdot {\bf x}'}$$|{\rm vac}\rangle$
$=e^{ {\bf x}\cdot{\bf x'}}$, we have $f(Q,S)$$=\{ Q, S|Q, S\}$
$=[T^{\ast Q}_{S}({\bf x})$$T^{Q}_{S}({\bf x}')$
$e^{{\bf x}\cdot {\bf x}'}]_{o}$
$=[T^{\ast Q}_{S}({\bf x})$$T^{Q}_{S}({\bf x}')$
$({\bf x}\cdot {\bf x}')^{2Q+S}]_{o}$$/(2Q+S)!$, which gives eq.(\ref{norm})
after differentiation. 

\bibitem{NS} P. Nozieres and D. Saint James, J. Phys. (Paris) {\bf 43}, 
1133 (1982), P. Nozieres in {\em Bose-Einstein Condensation}, ed. A. Griffin, 
D.W. Snoke, and S. Stringari (Cambridge University Press). 

\bibitem{CommentA} The correct normalization factor should include a factor 
$e^{\xi \sigma}$. However, this factor can be ignored as 
$\xi \sigma<<1$. 

\bibitem{LeeYang} K. Huang and C.N. Yang, Phys. Rev. {\bf 105}, 767 (1957). 

\bibitem{quadratic} For sufficiently large fields, the quadratic Zeeman effect
can force the fragmented state into a single condensate state with only 
$N_{0} \neq 0$. This way of recovering the single condensate state is rather 
trivial. We shall only consider the subtle case where the quadratic Zeeman
effect is weak. 

\bibitem{LDA}  These statements can be seen most
clearly in the local density / 
 Thomas-Fermi approximation. The ratio 
between $\lambda_{q}$ and the diagonal ``potential" term
is given by, for $1 << q << N_{-1}, N_{1}$;
$\sqrt{N_{-1} N_{1} } \int f_{0}^2 f_{1}f_{-1} \
/\int f^{2}_{0}[N_1 f^{2}_{1} + N_{-1} f^{2}_{-1}] \ < 1$.

\bibitem{crossover} The effect of quadratic Zeeman effect on the crossover
region will be studied elsewhere. 

\end{references}
\end{document}